\documentstyle[aps,multicol]{revtex}
\begin{document}
\draft

\title{Defect Relaxation and Coarsening Exponents}
\author{A. J. Bray}
\address{Department of Physics and Astronomy, The University, 
Manchester M13 9PL, UK}
\date{\today}
\maketitle

\begin{abstract}
The coarsening exponents describing the growth of long-range order in 
systems quenched from a disordered to an ordered phase are discussed 
in terms of the decay rate $\omega(k)$ for the relaxation of a 
distortion of wavevector $k$ applied to a topological defect.
For systems described by order parameters with $Z(2)$ (`Ising') and $O(2)$ 
(`XY') symmetry the appropriate defects are domain walls and vortex lines 
respectively. From $\omega_k \sim k^z$, we infer $L(t) \sim t^{1/z}$ for 
the coarsening scale, with the assumption that defect relaxation provides 
the dominant coarsening mechanism. The $O(2)$ case requires careful discussion
due to infrared divergences associated with the far field of a vortex line. 
Conserved, non-conserved, and `intermediate' dynamics are considered, with 
either short-range or long-range interactions. In all cases the results agree 
with an earlier `energy scaling' analysis. 

\end{abstract}

\begin{multicols}{2}

\narrowtext


\section{Introduction}
Coarsening is a characteristic property of a system quenched into an ordered 
phase from a disordered phase, and describes the establishment of long-range 
order over ever larger length scales as time increases \cite{Review}. We 
concentrate here on systems which exhibit the property of `dynamical scaling', 
in which the pair correlation function 
$C(r,t) \equiv \langle \phi({\bf x},t)\,\phi({\bf x}+{\bf r},t)\rangle$ 
of the order-parameter field $\phi({\bf r},t)$ has the scaling form 
$C(r,t) = f[r/L(t)]$. The characteristic length scale $L(t)$ generally has a 
power-law dependence on time, $L(t) \sim t^{1/z}$, sometimes with logarithmic 
corrections, where $z$ is a kind of dynamic exponent for the coarsening 
process.

Much effort in recent years has been devoted to the determination of the 
exponent $z$ (and any additional logarithmic factors) for a range of different 
models. Although the experimentally most relevant case of a scalar order 
parameter has been understood for some time, through intuitive arguments based 
on the scaling assumption, and through exactly solved limits \cite{LSW}, a 
general approach has been lacking until recently. The `energy-scaling' approach 
of Bray and Rutenberg \cite{Review,ES,RB} filled this gap by providing a very 
general approach based on the role of the topological defects nucleated during 
the quench. These defects are simple domain walls for a scalar field, 
vortices/vortex lines for a two-component vector field, and monopoles for 
three-component vector field (where we are assuming rotational invariance 
for the last two models, i.e.\ these models possess $O(2)$ and $O(3)$ symmetry 
respectively). The energy-scaling method is suitable for any system for which 
the scaling hypothesis holds and the dynamics are purely dissipative. These 
include the standard models `A' and `B' of the Hohenberg and Halperin 
classification \cite{HH}, describing the simplest systems with non-conserved 
and conserved order parameter respectively. In the context of phase-ordering 
kinetics, the models are described by the time-dependent Ginzburg-Landau 
(TDGL) equation (see section II) and the Cahn-Hilliard equation (section III).

Powerful though the energy-scaling method is, the arguments are not completely 
straightforward and some care is required in its application (see \cite{Review} 
for a detailed discussion). In this paper, therefore, we will discuss a much 
`cleaner' approach that addresses the dynamics of the topological defects in a 
more direct way. Specifically we consider a single plane domain wall, or 
straight vortex line, and apply a periodic perturbation of wavevector $k$. 
The perturbation then relaxes away, with asymptotic relaxation rate 
$\omega(k)$. A central assumption, as before, is the validity of the scaling 
hypothesis. If there is a single characteristic length scale $L(t)$, then the 
relaxation dynamics of the defect should be reflected in the coarsening 
dynamics, through $\dot{L}/L \sim \omega(k)$ with $k \sim 1/L$. So if 
$\omega(k) \sim k^z$ for $k \to 0$, we expect $L(t) \sim t^{1/z}$. 
More simply, one can set $\omega \sim 1/t$ and $k \sim 1/L$ to obtain the 
same result. 

This approach has been used before for systems with domain walls, and the 
conventional $z=2$ and $3$ for systems with conserved and nonconserved order 
parameter respectively have been recovered 
\cite{Langer,Jasnow,Shinozaki1}. Here we generalize to systems with 
`intermediate' dynamics (to be defined below), and/or long-range interactions, 
and extend the method to systems with $O(2)$ symmetry. In this way we can 
confront most of the predictions of the energy scaling method. Since the 
present approach requires extended defects, however, it cannot treat the 
$O(2)$ system in $d=2$ spatial dimensions, or the $O(3)$ system in $d=3$, 
for which the relevant topological defects (vortices and monopoles 
respectively) are points. We therefore restrict our attention to scalar 
systems in $d=2,3$ and $O(2)$ systems in $d=3$. For these cases, all our 
results agree with the energy scaling predictions, although some subtleties 
arise in the case of the $O(2)$ system due to infrared singularities 
associated with the far field of the vortex line.

The paper is organized as follows. Section II deals with nonconserved 
dynamics (`model A'). The method is introduced in the simple context of 
domain walls. The extension to vortex lines is then discussed, followed by the 
generalization to long-range interactions. Section III deals with conserved 
(`model B') and `partially conserved' dynamics, for domain walls and vortex 
lines with both and short-range interactions. Detailed comparisons with the 
predictions of the energy scaling method are made at each stage. In particular, 
the energy scaling prediction  $L(t) \sim (t\ln t)^{1/4}$ for the conserved 
$O(2)$ model in $d=3$ (or, more generally, $d \ge 3$) is recovered. The paper 
concludes with a discussion and summary.

\section{Nonconserved Dynamics}
\label{Nonconserved}
\subsection{Domain walls}
\label{walls}
The simplest exemplar of the defect relaxation approach is the dynamics 
of an interface (domain wall) separating two equilibrium phases. 
The time-dependent Ginzburg Landau (TDGL) equation is a continuum dynamical 
model for such a process:
\begin{equation}
\partial_t \phi = \nabla^2 \phi - V'(\phi)\ ,
\label{TDGL}
\end{equation}
where $\phi({\bf x},t)$ is the scalar order-parameter field and $V(\phi)$ 
is a symmetric double-well potential, e.g.\ $V(\phi)=(1-\phi^2)^2/4$, whose 
minima $\phi=\pm 1$ represent the two bulk phases.

Consider first a single planar interface separating the two phases. 
The normal to the interface define the $z$-direction. The order parameter 
depends only on $z$, and satisfies the time-independent version of 
(\ref{TDGL}), $d^2\phi_0/dz^2 = V'(\phi_0)$, with boundary conditions 
$\phi_0(\pm \infty) = \pm 1$. We now impose a small periodic perturbation 
to the interface in the $x$-direction with wavevector $k$ 
\begin{equation}
\phi(x,z,t) = \phi_0(z) + A\phi_1(z)\exp(ikx - \omega_k t)\ ,
\label{mod}
\end{equation}
where the amplitude $A$ is small. The relaxation rate $\omega_k$ defines the 
timescale, $\tau_k = 1/\omega_k$, for the relaxation of a perturbation 
with characteristic length-scale $2\pi/k$. Substituting this form into 
(\ref{TDGL}), and linearizing in $A$, gives the eigenvalue equation 
$H\phi_1 = \omega_k \phi_1$, with `Hamiltonian'
\begin{equation}
H = - \frac{d^2}{dz^2} + k^2 + V''(\phi_0)\ .
\end{equation}
This is conveniently thought of as a quantum mechanical Hamiltonian 
operator. Note that $V''(\phi_0)$ is positive at $z = \pm \infty$ 
(where $\phi_0 =\pm 1$), and negative at $z=0$ (where $\phi_0=0$). 
For the specific case $V(\phi)=(1-\phi^2)^2/4$ one has 
$V''(\phi_0) = 3\phi_0^2(z)-1$, which equals $-1$ at $z=0$ and tends 
to $+2$ for $z \to \pm \infty$. It follows that $V''(\phi_0)$ represents 
a potential well which must have at least one bound state. In fact, 
since $k=0$ corresponds to a uniform translation of the interface,  we know 
that $\omega_{k=0}=0$, with eigenfunction $\phi_1=d\phi_0/dz$. Also, 
since this function has no nodes, it must be the ground state. 
Since the $k$-dependence of $H$ is simply the additive constant $k^2$ in this 
simple model, it follows that $d\phi_0/dz$ is the ground state for all $k$, 
with eigenvalue 
\begin{equation}
\omega_k = k^2. 
\label{disA}
\end{equation}
Higher eigenvalues are separated by a gap from the ground state, 
so any component of the corresponding eigenfunctions, in the initial 
displacement of the interface, relaxes quickly to zero. For this reason, 
we concentrate here only on the groundstate eigenfunction, which is the 
`slow mode'. Neglecting any contribution from the other eigenfunctions, 
(\ref{mod}) can be written, correct to leading order in the amplitude $A$, 
as $\phi(z,x,t) = \phi_0(z+A\exp[ikx - \omega_k t])$, corresponding simply 
to a modulated interface located at $z(x,t) = - A \exp[ikx - \omega_k t]$. 
Note that this simple interpretation is a consequence of $\phi = d\phi_0/dz$, 
which will not be true in general.

Equation (\ref{disA}) has the form $\omega_k = k^z$ with $z=2$, from which 
infer the coarsening growth law $L(t) \sim t^{1/z} = t^{1/2}$ for this, the 
simplest of our dynamical models.

\subsection{Vortex Lines}
\label{vortex}
If the order-parameter field is the two-component vector field 
$\vec{\phi}({\bf x},t)$ of the $O(2)$ model, the topological defects are 
vortex lines. The analogue of (\ref{TDGL}) for this system is 
\begin{equation}
\partial_t \vec{\phi} = \nabla^2 \vec{\phi} - dV/d\vec{\phi}\ ,
\label{TDGL2}
\end{equation}
where the `wine-bottle' potential $V(\vec{\phi})$ is a function of 
$|\vec{\phi}|$ only. For definiteness, we will use the form $V(\vec{\phi}) = 
(1 - |\vec{\phi}|^2)^2/4$, with minima on the ground-state manifold 
$|\vec{\phi}|=1$, although the conclusions are completely general. For 
this potential (\ref{TDGL2}) becomes
\begin{equation}
\partial_t \vec{\phi} = \nabla^2 \vec{\phi} 
+ (1 - |\vec{\phi}|^2)\,\vec{\phi}\ . 
\label{TDGL2a}
\end{equation}
This equation has a time-independent solution $\vec{\phi}_0({\bf r})$, 
corresponding to a vortex line in the $z$ direction, where we have 
introduced the coordinate system ${\bf x} = ({\bf r},z)$ of coordinates 
normal and parallel to the vortex. This is a solution of (\ref{TDGL2a}) 
satisfying the boundary conditions $\vec{\phi}_0(0)=0$, 
$\vec{\phi}_0({\bf r}) \to \hat{\bf r}$ for $|{\bf r}| \to \infty$, where 
$\hat{\bf r}$ is a unit vector.

By analogy with (\ref{mod}) we now perturb the vortex line by adding 
a periodic modulation with wavevector $k$ in the form 
\begin{equation}
\vec{\phi}({\bf r},z,t) = \vec{\phi}_0({\bf r}) 
+ A\vec{\phi}_1({\bf r})\exp(ikz - \omega_k t)\ .
\label{mod2}
\end{equation}
Substituting in (\ref{TDGL2a}), and linearizing in A, gives the eigenvalue 
equation 
\begin{equation}
(\nabla_r^2 - k^2)\vec{\phi}_1 + (1 - |\vec{\phi}_0|^2)\,\vec{\phi}_1 
 - 2 (\vec{\phi}_1\cdot\vec{\phi}_0)\,\vec{\phi}_0 = -\omega_k \vec{\phi}_1\ ,
\label{eigen2}
\end{equation}
where $\nabla_r^2$ is the Laplacian operator in the ${\bf r}$-plane. 
From the same physical considerations that we used for domain walls, 
we expect a null eigenfunction at $k=0$, corresponding to a uniform 
displacement of the vortex line transverse to its length. Since 
$\vec{\phi}_0({\bf r} + {\bf a}) = \vec{\phi}_0 
+ {\bf a}\cdot\nabla_r \vec{\phi}_0$, to first order in ${\bf a}$, we 
identify a family of null eigenfunctions 
$\vec{\phi}_1 = {\bf a}\cdot\nabla_r \vec{\phi}_0$, parametrized by the 
direction of ${\bf a}$. For convenience we may choose a basis set of 
two such eigenfunctions corresponding to ${\bf a}$ being a unit vector 
along the $x$ and $y$ axes (where ${\bf r} = (x,y)$) respectively. The 
two basis functions are then $\partial_x \vec{\phi}_0$ and 
$\partial_y \vec{\phi}_0$. They are orthogonal by symmetry. 
Substitution into (\ref{eigen2}) confirms that 
these are null eigenfunctions, i.e.\ $\omega_{\bf k=0}=0$. Since the 
$k^2$ term in (\ref{eigen2}) can simply be absorbed into $\omega_k$, 
it follows that, just as for domain walls, the smallest eigenvalue  
for any $k$ is exactly given by
\begin{equation}
\omega_k = k^2\ .
\label{dis2A}
\end{equation}
We conclude that the dynamical exponent is again $z=2$. This agrees with 
the energy-scaling result.

For future reference we note that the eigenfunctions 
$\partial_x \vec{\phi}_0$ and $\partial_y \vec{\phi}_0$ are not normalizable 
in an infinite two-dimensional space,  since the normalization integrals 
are logarithmically divergent (see below). When the normalization is 
important, we will take the vortex line to lie along the symmetry axis of 
an infinitely long cylinder of radius $R$.  

%DISCUSSION: (i) Only d=3, not d=3 (ii) No `gap', since can modulate the 
%direction of the displacement direction a at some slow rate q as one goes 
%down the vortex line. Then one would expect a relaxation rate of order 
%$(k^2 + q^2)$? Corrections to scaling more important than for walls? 

\subsection{Long-Range Interactions: Domain Walls}
The above approach may readily be generalized to the case where the 
underlying interactions are long-ranged in space. Consider, for example,  
a ferromagnetic model, in which the exchange interaction 
$J(|{\bf r}-{\bf r'}|)$ falls off with distance as 
$|{\bf r}-{\bf r'}|^{-(d+\sigma)}$ for large $|{\bf r}-{\bf r'}|$. The 
long-range character of these interactions is typically only `relevant' 
(i.e.\ affects the scaling behavior) if $\sigma$ is smaller than a critical 
value $\sigma_c$ \cite{Bray93}. In the present context, we will find that 
$\sigma_c \le 2$, so we will consider only the case $\sigma < 2$ here. 
For this case, the term in the Ginzburg-Landau free-energy functional 
generated by long-range interactions has the Fourier-space form 
$F_{LR} = (g/2) \sum_{\bf k} k^\sigma \phi_{\bf k}\phi_{-\bf k}$.  
This can be formally represented in the TDGL dynamics, 
$\partial_t \phi = -\delta F/\delta \phi$, by replacing the usual 
Laplacian operator by $ -(-\nabla^2)^{\sigma/2}$, to give 
\begin{equation} 
\partial_t \phi = \nabla^2\phi - g(-\nabla^2)^{\sigma/2} \phi - V'(\phi)\ ,
\label{TDGL-LR}
\end{equation}
instead of (\ref{TDGL}). Expanding around a flat domain wall using 
(\ref{mod}), where $\phi_0$ is now a stationary solution of the full 
equation (\ref{TDGL-LR}), gives the eigenvalue equation 
\begin{equation}
(d_z^2 - k^2 - g[k^2 - d_z^2]^{\sigma/2} - V''[\phi_0])\phi_1 = 
-\omega_k \phi_1\ ,
\label{eigen_LR}
\end{equation}
where $d_z \equiv d/dz$.

The function $\phi_1=d\phi_0/dz$ is again a null eigenfunction for $k=0$, 
since $k=0$ corresponds to a simple translation of the interface. The term 
in $k^2$ in (\ref{eigen_LR}) simply shifts the eigenvalue by $k^2$, as 
before. However, the long-range part of (\ref{eigen_LR}) modifies the 
eigenfunction. We can, nevertheless, compute the desired small-$k$ 
behavior of the ground-state eigenvalue perturbatively, using the 
unperturbed ground state eigenfunction $d\phi_0/dz \equiv \phi_0'(z)$: 
\begin{equation}
\omega_k = k^2 + g\,\frac{\int_{-\infty}^\infty dz\,\phi_0'(z)\, 
([k^2 - d_z^2]^{\sigma/2} - [-d_z^2]^{\sigma/2})\,\phi_0'(z)}
{\int_{-\infty}^\infty dz\,[\phi_0'(z)]^2}\ .
\label{dis-LR}
\end{equation}
The integral in the numerator is conveniently evaluated in Fourier space. 
The function $\phi_0'(z)$ is sharply peaked at the interface, with a width 
of order $\xi$, the interfacial thickness, and a peak height of order 
$1/\xi$. Its Fourier transform, therefore, is very broad (with width $1/\xi$) 
and for $q\xi \ll 1$ is equal to  $\phi_0(\infty) - \phi_0(-\infty) = 2$.   
The integral in the numerator in (\ref{dis-LR}) becomes, therefore 
$$
4\int_{-\infty}^\infty \frac{dq}{2\pi}([k^2+q^2]^{\sigma/2} - |q|^\sigma),
$$
for small $k$, provided the integral converges. The latter condition 
requires $\sigma < 1$. In this regime, the important values of $q$ in the 
integral are of order $k$, so our replacement of $\phi_0'(z)$ by 
$2\delta(z)$ is justified for $k\xi \ll 1$. The integral is then easily 
evaluated to give $k^{1+\sigma}$ (up to constants) for $\sigma < 1$.

For $\sigma>1$, the replacement of $\phi_0'$ by a delta function is no 
longer valid, since the integral in the numerator in (\ref{dis-LR}) 
would not converge. The Fourier transform of $\phi_0'$ falls off at 
$q\xi \approx 1$, however, to converge the integral. In this case 
($\sigma >1$) the characteristic value of $q$ is of order $1/\xi$, 
instead of $k$, and we can expand the integrand up to order $k^2$ for 
$k\xi \ll 1$. The numerator then becomes of order $k^2 \xi^{1-\sigma}$ for 
$\sigma >1$. A similar line of reasoning for the marginal case $\sigma=1$ 
leads to a $k^2\ln(1/k\xi)$ behavior for $k\xi \ll 1$. The simple $k^2$ 
dependence for $\sigma >1$ just renormalizes the amplitude of the leading 
$k^2$ term in (\ref{dis-LR}). The integral in the denominator in 
(\ref{dis-LR}) is a constant of order $1/\xi$.

Putting it all together, retaining only the leading small-$k$ behavior, 
and discarding constant prefactors, gives the dispersion relation 
\begin{eqnarray}
\omega_k & \sim & k^{1+\sigma},\ \ \ \ \sigma<1, 
\nonumber \\
 & \sim & k^2\,\ln(1/k\xi),\ \ \ \ \sigma=1, \nonumber \\
 & \sim & k^2,\ \ \ \ \sigma>1.
\label{disA-LR}
\end{eqnarray} 
We deduce that the dynamic exponent is $z=1+\sigma$ for $\sigma<1$ and 
$z=2$ for $\sigma>1$. The marginal case $\sigma=1$ gives (with $k \to 1/L$, 
$\omega_k \to 1/t$ as usual) the growth law $L \sim (t\ln t)^{1/2}$. All 
these results are in accord with previous results based on renormalization 
group \cite{Bray93} and energy scaling \cite{ES,RB} arguments.

\subsection{Long-Range Interactions: Vortex Lines}
The influence of long-range interactions on the dynamics of a vortex line 
may be discussed in a similar way. Taking, as before, $V(\vec{\phi}) = 
(1-|\vec{\phi}|^2)^2/4$, one obtains, analogous to (\ref{eigen2}) and 
(\ref{eigen_LR}),
\begin{eqnarray}
(\nabla_r^2 - k^2 - g[k^2 - \nabla_r^2]^{\sigma/2} 
+ 1 - |\vec{\phi}_0|^2)\vec{\phi}_1 & \nonumber \\
- 2 (\vec{\phi}_1\cdot\vec{\phi}_0)\vec{\phi}_0 = -\omega_k \vec{\phi}_1\ .&
\label{eigen_LR2}
\end{eqnarray}
The long-range part can be treated perturbatively for small $k$ as in the 
scalar case. Since the perturbation is isotropic, the unperturbed  
eigenfunctions $\partial_x\vec{\phi}_0$ and $\partial_y\vec{\phi}_0$ are 
not mixed by the perturbation. It follows that, to leading order 
in perturbation theory, 
\begin{equation}
\omega_k = k^2 + g\,\frac{\int d^2r\,\partial_i\phi_{0j}\, 
([k^2 - \nabla_r^2]^{\sigma/2} - [-\nabla_r^2]^{\sigma/2})\,
\partial_i\phi_{0j}}
{\int d^2r\,[\partial_i\phi_{0j}]^2}\ ,
\label{dis2-LR}
\end{equation}
where the result has been written in a rotationally invariant form, 
i.e.\ there are implicit summations over the indices $i$ and $j$.

We again evaluate the integral in the numerator in Fourier space. In the 
small-$k$ limit only small wavevector ${\bf q}$, corresponding to large 
$|{\bf r}|$, will be important. For large $|{\bf r}|$, 
$\vec{\phi}_0 \to \hat{\bf r}$, giving $\partial_i\phi_{0j} = (\delta_{ij} 
- \hat{r}_i\hat{r}_j)/|{\bf r}|$. The Fourier transform of this quantity is 
$(2\pi/q)(\delta_{ij} - \hat{q}_i\hat{q}_j)$. Inserting this in 
(\ref{dis2-LR}) the integral in the numerator becomes, up to constants, 
\begin{equation}
\int_{|{\bf q}|>1/R} \frac{d^2q}{q^2}([k^2 + q^2]^{\sigma/2} 
- |{\bf q}|^\sigma)\ \sim k^\sigma\,\ln(kR)
\label{num}
\end{equation}
for $\sigma < 2$, this condition ensuring the convergence of the 
integral at large $q$. The lower cutoff at $q=1/R$, where we recall that 
$R$ is the radius of the system in the plane normal to the vortex line, is 
necessary to regulate the logarithmic singularity at small $q$.

Now consider the integral in the denominator of (\ref{dis2-LR}). Using the 
large-$r$ form of $\vec{\phi}_0$ everywhere gives the result  
$\int d^2r/r^2$, i.e.\ a logarithmically divergent integral. At small $r$, 
this can be cut-off at the vortex core size $\xi$, where the assumed 
large-$r$ form no longer holds. The large $r$ cut-off is again $R$. 
The need for a large-distance cut-off is associated with the 
non-normalizability of the null eigenfunctions, as discussed in section 
\ref{vortex}. The upshot is that the denominator is of order $\ln(R/\xi)$, 
and the leading small-$k$ form of the dispersion relation is 
\begin{equation}
\omega_k \sim k^\sigma\,\frac{\ln(kR)}{\ln(R/\xi)},\ \ \ \ \sigma < 2.
\label{dis2A-LR}
\end{equation}

Some discussion of this result is in order. If we take the limit 
$R \to \infty$ in (\ref{dis2A-LR}), we obtain the well-defined limit 
$\omega_k \sim k^\sigma$ for a vortex line in an infinite system. 
We argue, however, that this is not the appropriate limit in which to 
discuss the implications of (\ref{dis2A-LR}) for the coarsening dynamics. 
In a system with many vortex lines, a characteristic scale $L$ can be 
associated with the line density $\rho_V$ ($=$ length of vortex line per 
unit volume) via $\rho_V=1/L^2$. Since the far field of a given vortex 
line is screened out on this scale, it is $L$ rather than $R$ which is 
the appropriate cut-off in the coarsening system. Replacing $R$ by $L$ 
in (\ref{dis2A-LR}), and making the usual identifications $k \sim 1/L$ 
and $\omega_k \sim 1/t$, gives the coarsening growth law 
$L \sim (t/\ln t)^{1/\sigma}$. This agrees again with the predictions 
of the energy-scaling approach. Similar arguments concerning the nature of 
the momentum cutoffs will be necessary for the conserved $0(2)$ model 
in section \ref{vortexB}.

\section{Conserved Dynamics}
\label{Conserved}
\subsection{Domain Walls}
\label{wallsB}
The standard continuum model for the time evolution of a conserved scalar 
field is the Cahn-Hilliard equation 
\begin{equation}
\partial_t\phi= -\nabla^2 [\nabla^2 \phi - V'(\phi)]\ .
\end{equation}
For present purposes this may be conveniently rewritten in the form 
\begin{equation}
-\nabla^2 \phi + V'(\phi) + (-\nabla^2)^{-1} \partial_t\phi = 0\ .
\end{equation}
Inserting the form (\ref{mod}) and linearizing in $A$ leads to the 
eigenvalue equation 
\begin{eqnarray}
[k^2 \phi_1(z) - \omega_k \int_{-\infty}^\infty dz'\,G_k(z-z')\phi_1(z')]&
\nonumber \\
+ [-d_z^2 + V''(\phi_0)]\phi_1(z) =0\ ,& 
\label{eigenB}
\end{eqnarray}
where 
\begin{eqnarray}
G_k(z-z') &=& \int_{-\infty}^\infty \frac{dq}{2\pi}\, 
\frac{\exp[iq(z-z')]}{k^2 + q^2} \nonumber \\ 
 &=& \frac{\exp(-k|z-z'|)}{2k} 
\label{Green's}
\end{eqnarray}
is the Green's function for the operator $(k^2-d_z^2)$.

As before, $\phi_1 = \phi_0'$ is a null eigenfunction for $k=0$. For small 
$k$ the term in the second square bracket in (\ref{eigenB}), which vanishes 
for $k=0$, can be treated perturbatively to give 
\begin{equation}
\omega_k = k^2\,\frac{\int_{-\infty}^\infty dz\,[\phi_0'(z)]^2}
{\int_{-\infty}^\infty dz \int_{-\infty}^\infty dz'\,
G_k(z-z')\,\phi_0'(z)\,\phi_0'(z')}\ .
\label{dispB}
\end{equation}
The integral in the numerator defines the surface tension, 
$\gamma = \int_{-\infty}^\infty dz\,(\phi_0')^2 \approx 1/\xi$. The 
function $\phi_0'(z)$ acts like a smeared delta function of width $\xi$ and 
strength 2. For $k\xi \ll 1$, therefore, we can replace $G_k(z-z')$ by its 
small-argument limit $1/2k$ in the denominator, to give the result $2/k$. 
In this limit, therefore,
\begin{equation}
\omega_k = \frac{1}{2}\,\gamma k^3\ .
\label{disB}
\end{equation}
The dynamic exponent is $z=3$, in accord with the expected Lifshitz-Slyozov 
scaling $L \sim t^{1/3}$. A more careful derivation of this result has been 
given by Shinozaki and Oono \cite{Shinozaki1}.

The generalization to `intermediate' dynamics, or `noninteger derivative 
models' \cite{Onuki}, in which the leading $-\nabla^2$ in the 
Cahn-Hilliard equation is replaced by $(-\nabla^2)^{\mu/2}$ 
(i.e.\ by $|{\bf k}|^\mu$ in Fourier space, instead of $k^2$) 
is straightforward. One simply has to insert in (\ref{dispB}) 
the Green's function for $(k^2 - d_z^2)^{\mu/2}$, given by 
\begin{equation}
G_k(z-z') = \int_{-\infty}^\infty \frac{dq}{2\pi}\,\frac{\exp[iq(z-z')]}
{(k^2 + q^2)^{\mu/2}}\ .
\label{Green-mu}
\end{equation}
For $\mu=2$ our previous result is obtained, while for $\mu=0$ one gets 
$G_k(z-z') = \delta(z-z')$, giving the nonconserved result $\omega_k = k^2$.
For general $\mu$, (\ref{Green-mu}) gives the scaling form 
$G_k(z-z') = k^{1-\mu}f(k|z-z'|)$. From the discussion of the case $\mu=2$ 
it is clear that we require the result only in the limit $k|z-z'| \ll 1$, 
i.e.\ we need the small-argument form of the scaling function $f(x)$. 
Straightforward analysis gives $f(0) = {\rm const.}$ for $\mu >1$, 
and $f(x) \sim x^{\mu -1}$ for $\mu < 1$, with $f(x) \sim \ln(1/x)$ for 
$\mu=1$. Thus there is a change of behavior at $\mu=1$. For $\mu>1$, 
$G_k(z-z') \sim k^{1-\mu}$ gives (with $\gamma \approx 1/\xi$) \ 
$\omega_k \sim \xi^{-1} k^{1+\mu}$. For $\mu <1$, the double integral 
$\int dz \int dz'\,|z-z'|^{\mu-1}\phi_0'(z)\phi_0(z')$ 
is of order $\xi^{\mu-1}$, giving the dispersion relation 
$\omega_k \sim \xi^{-\mu}k^2$, while for $\mu=1$ one obtains 
$\omega_k \sim \xi^{-1}k^2/\ln(1/k\xi)$. For convenience we summarize 
these results below, dropping prefactors involving $\xi$:
\begin{eqnarray}
\omega_k & \sim & k^{1+\mu},\ \ \ \ \mu>1 \nonumber \\
         & \sim & \frac{k^2}{\ln (1/k\xi)},\ \ \ \ \mu=1 \nonumber \\
         & \sim & k^2, \ \ \ \ \mu <1\ .
\end{eqnarray}		 
These results have been derived previously by Onuki \cite{Onuki}. 
The dynamic exponent is $z=1+\mu$ for $\mu>1$ and $z=2$ for $\mu<1$. 
For $\mu < 1$, the exponent is the same as for the nonconserved 
system: the conservation law is `irrelevant', in accord with a  
renormalization group argument \cite{Bray90}.

Using $\omega \sim 1/t$ and $k \sim 1/L$ gives the growth laws 
$L(t) \sim t^{1/(1+\mu)}$ for $\mu >1$, $(t/\ln t)^{1/2}$ for $\mu=1$, 
and the usual nonconserved result $L(t) \sim t^{1/2}$ for $\mu<1$. 
These results are again in complete accord with the energy-scaling 
argument. 

Long-range interactions can be included for a conserved order parameter 
in a straightforward way. Omitting the details, the final 
result is a rather obvious combination of (\ref{dis-LR}) and 
(\ref{dispB}). For general $\mu$ it reads
\begin{equation}
\omega_k = \frac{\int_{-\infty}^\infty dz\,\phi_0'(z)\,O_k\,\phi_0'(z)}
{\int_{-\infty}^\infty dz \int_{-\infty}^\infty dz'\,
G_k(z-z')\,\phi_0'(z)\,\phi_0'(z')}\ ,
\end{equation}
where $O_k$ is the operator
\begin{equation}
O_k = k^2 + g[(k^2-d_z^2)^{\sigma/2}-(-d_z^2)^{\sigma/2}].
\end{equation}
Note that the numerator and denominator contain the information about the 
interactions and the conservation law respectively. It follows that the 
dynamic exponent is $z(\sigma,\mu) = \min(1+\sigma,2) + \max(\mu-1,0)$. 
This clean separation of the role the interactions and the conservation 
law is mirrored in the `energy-scaling' approach to calculating growth 
exponents, where the energy depends only on the interactions ($\sigma$) 
and the energy dissipation rate only on the conservation law ($\mu$).

\subsection{Vortex Lines}
\label{vortexB}
The treatment of vortex lines in a conserved $0(2)$ model follows the same 
pattern. Consider first the case of `simple' conservation, $\mu=2$. 
The starting point is the Cahn-Hilliard equation for vector fields:
\begin{equation}
(-\nabla^2)^{-1} \partial_t \vec{\phi} = \nabla^2 \vec{\phi} + 
(1 - |\vec{\phi}|^2) \vec{\phi}\ .
\end{equation}
Expanding around the stationary vortex solution $\vec{\phi}_0$ using 
(\ref{mod2}) gives the eigenvalue equation
\begin{eqnarray}
\left[-\nabla_r^2 - \left(1-|\vec{\phi}_0({\bf r})|^2\right)\right]\,
\vec{\phi}_1({\bf r}) 
+ 2\left(\vec{\phi}_0({\bf r})\cdot\vec{\phi}_1({\bf r})\right)\,
\vec{\phi}_0 & \nonumber \\ 
+ \left[k^2 - \omega_k \int d^2r'\,G_k({\bf r} - {\bf r'})\,
\vec{\phi}_1({\bf r'})\right]=0\ ,& 
\end{eqnarray}
where 
\begin{equation}
G_k({\bf r} - {\bf r'}) = \int \frac{d^2q}{(2\pi)^2}\,
\frac{\exp(i{\bf q}\cdot{\bf r})}{k^2+q^2} 
\label{Green's2}
\end{equation}
is the Green's function for $(k^2-\nabla_r^2)$.

For small $k$, the eigenvalue $\omega_k$ can again be calculated using 
first-order perturbation theory. Rotational invariance ensures, as before, 
that the unperturbed eigenfunctions $\partial_i\phi_{0j}$ are not mixed by 
the perturbation. The result can be written in the form 
\begin{eqnarray}
\omega_k & = & k^2\,\frac{\int d^2r\,(\nabla \vec{\phi}_0)^2}
{\int d^2r \int d^2r'\,G_k(|{\bf r}-{\bf r}'|)\,\nabla \vec{\phi}_0({\bf r})
\cdot\nabla \vec{\phi}_0({\bf r}')} \nonumber \\
 & = & k^2\,\frac{\int d^2q\, q^2\, |\vec{\phi}_0({\bf q})|^2}{\int d^2q\,
 [q^2/(k^2+q^2)]\,|\vec{\phi}_0({\bf q})|^2}\ .
\label{disp2B}
\end{eqnarray}
The Fourier-space form is more convenient for present purposes. From the 
result $\vec{\phi}_0 \to \hat{\bf r}$ for $r \gg \xi$ it follows that 
$|\vec{\phi}_0({\bf q})|^2 \to (2\pi)^2/q^4$ for $q\xi \ll 1$. 
Using this result for all $q$, and introducing ultraviolet and infrared 
cutoffs $1/\xi$ and $1/R$ respectively as required, the result for 
$\omega_k$ can be written as
\begin{eqnarray}
\omega_k & = & k^2\,\frac{\int^{1/\xi}_{1/R} q\,dq/q^2}{\int^\infty_{1/R} 
q\,dq/[q^2(k^2 + q^2)]} \nonumber \\
 & = & k^4\,\frac{\ln(R/\xi)}{\ln(kR)}
\label{dis2B}
\end{eqnarray}
to leading logarithmic accuracy.

One again this result requires careful interpretation. Taking the limit 
$R \to \infty$ at fixed $k$ gives $\omega_k = k^4$, suggesting the 
coarsening growth law $L(t) \sim t^{1/4}$. The $R$-dependence enters 
(\ref{dis2B}), however, from the requirement to cut off infrared divergences 
associated with the far field of the vortex. In the phase-ordering 
context, the far field is cut off at scale $L$ by other vortex lines. 
This means we should replace $R$ by $L$ in (\ref{dis2B}) (see the 
parallel discussion after (\ref{dis2A-LR})). With $k \sim 1/L$ and 
$\omega_k \sim 1/t$ as usual, this leads to the coarsening growth law
$L(t) \sim (t\ln t)^{1/4}$, in perfect agreement with the energy-scaling 
result \cite{ES,RB}.

The extension to intermediate dynamics, controlled by an exponent $\mu$, 
is again straightforward. We simply replace $(k^2+q^2)$ by 
$(k^2+q^2)^{\mu/2}$ in (\ref{dis2B}), with the result
\begin{equation}
\omega_k = k^{2+\mu}\,\frac{\ln(R/\xi)}{\ln(kR)}\ ,
\end{equation}
to leading logarithmic accuracy, for any $\mu>0$. The coarsening growth 
law becomes $L \sim (t\ln t)^{1/(2+\mu)}$.

Finally, long-range interactions can be included. The expression for 
$\omega_k$ combines the numerator from (\ref{dis2-LR}) with the denominator 
from (\ref{disp2B}). Replacing $(k^2+q^2)$ by $(k^2+q^2)^{\mu/2}$ in 
the denominator, and using (\ref{num}) for the numerator, gives the result 
for general $\mu$ and $\sigma <2$:
\begin{equation}
\omega_k \sim k^{\sigma + \mu}\ .
\label{dis2B-LR}
\end{equation}
Strictly the numerator and denominator generate factors $\ln(\alpha kR)$ 
and $\ln(\beta kR)$ (with $\alpha \ne \beta$ in general) which we have 
cancelled in (\ref{dis2B-LR}). The cancellation is strictly valid in the 
limit $R \to \infty$ at fixed $k$. In the coarsening context, where  
$k \sim 1/L$ and $R \sim L$, these factors are of order unity so 
(\ref{dis2B-LR}) still holds, to give $z=\sigma + \mu$.

\section{Conclusion}
We have discussed a general method, based on the relaxation of a 
sinusoidally perturbed topological defect (e.g.\ a domain wall or vortex 
line), for inferring growth laws in phase-ordering systems. The underlying 
assumption is that there is a single characteristic length scale (`dynamical 
scaling') so that defect relaxation is either the sole or dominant coarsening 
mechanism, or occurs at the same rate as the underlying coarsening process. 
If this assumption does not hold, there need be no connection between 
the relaxation spectrum and the coarsening exponent. An example where the 
present approach fails has recently been given \cite{Siegert}. 

In this paper we have considered only systems with purely dissipative 
dynamics. These systems can also be addressed using energy-scaling 
arguments, which involve equating two independent estimates of the 
energy dissipation rate \cite{ES,RB}. In all cases, the results from the 
two different approaches agree. As yet, however, it has not proved possible 
to extend the energy-scaling technique beyond purely dissipative systems. 
The defect relaxation method, however, does not suffer from this 
limitation. Indeed, Shinozaki \cite{Shinozaki2} has studied the  
interfacial relaxation spectrum in an incompressible binary fluid, 
including hydrodynamic effects, and has obtained $\omega_k \sim k$ for  
$k \to 0$, consistent with the linear growth, $L(t) \sim t$, of the 
coarsening scale predicted by Siggia \cite{Siggia} for bicontinuous phases. 

Another recent application of this approach is to the coarsening of 
systems which exhibit lamellar structures in equilibrium, such as 
Rayleigh-B\'enard convective rolls as described by the Swift-Hohenberg 
equation, or block copolymers in the weak segregation regime \cite{Jacob}.
The dynamics of these systems has attracted much recent attention 
\cite{Lamellar}. 

In the coarsening regime the lamellae do not form parallel stripes but 
rather exhibit a globally isotropic, but locally striped, structure with a 
characteristic length scale (as measured, for example, by the typical 
radius of curvature of the stripes) which increases with time, and gives 
the length scale over which the stripes are locally roughly parallel. 
The stripe pattern itself has, in the two-dimensional systems studied 
numerically, topological defects in the form of disclinations. It is not 
yet clear whether the coarsening is described by a single growing length 
scale in these systems, as different measures of this scale give 
different results \cite{Lamellar}. An analysis of the dynamics of a 
modulated lamellar structure along the lines presented here gives the
relaxation rate  $\omega_k = (\epsilon/256)k^2 + k^4$, where $\epsilon$ 
is measure of the quench depth \cite{Jacob}. For shallow quenches, therefore, 
one has $\omega_k \simeq k^4$, suggesting a $t^{1/4}$ growth at not-too-late 
times, whereas the growth rate inferred from the evolution of the structure 
factor is closer to $t^{1/5}$ \cite{Lamellar}. The disclinations, however, 
have not been included in this approach, and it is possible that these 
coarsen more slowly than the interfaces. More work is needed to clarify 
this point.    

In summary, a study of defect relaxation provides a simple way to 
determine growth exponents in coarsening systems. This approach should 
be reliable when the coarsening is described by a single characteristic 
length scale. In the cases studied here, the results obtained are 
identical to those derived using the energy-scaling method. While the 
underlying assumptions (dynamical scaling) are the same for both methods, 
the ideas involved in the present approach are rather simpler, and the 
method is not restricted to purely dissipative dynamics. The energy-scaling 
method, on the other hand, is not restricted to extended defects, i.e.\ 
it can be used for systems with point defects or, indeed, no defects at all.   

\begin{small}
\begin{center}
{\bf ACKNOWLEDGEMENT}  
\end{center}
\end{small}
The author thanks Jacob Christensen and Thomas Prellberg for discussions. 
This work was supported by the Engineering and Physical Sciences Research 
Council (UK).

\end{multicols}

\end{document}